\documentclass{elsart}
\usepackage{graphicx}

\begin{document}

\begin{frontmatter}

\title{Applying Shower Development Universality to KASCADE Data}

\author[fzk]{W.D. Apel},
\author[fzk]{A.F. Badea\thanksref{r1}}, 
\author[fzk]{K. Bekk},
\author[fzk,uni]{J. Bl\"umer},
\author[kas]{E. Boos},
\author[fzk]{H. Bozdog},
\author[rum]{I.M. Brancus}, 
\author[fzk]{K. Daumiller},
\author[fzk]{P. Doll},
\author[fzk]{R. Engel}, 
\author[fzk]{J. Engler}, 
\author[fzk]{H.J. Gils},
\author[wup]{R. Glasstetter}, 
\author[fzk]{A. Haungs\thanksref{corr}},
\author[fzk]{D. Heck},
\author[uni]{J.R. H\"orandel\thanksref{now}}, 
\author[wup]{K.-H. Kampert}, 
\author[fzk]{H.O. Klages},
\author[fzk,kas]{I. Lebedev\thanksref{corr}},
\author[fzk]{H.J. Mathes}, 
\author[fzk]{H.J. Mayer},
\author[fzk]{J. Milke},
\author[fzk]{J. Oehlschl\"ager}, 
\author[fzk]{S. Ostapchenko\thanksref{r4}}, 
\author[rum]{M. Petcu},
\author[fzk]{H. Rebel},
\author[fzk]{M. Roth},
\author[fzk]{G. Schatz},
\author[fzk]{H. Schieler}, 
\author[fzk]{H. Ulrich},
\author[fzk]{J. van Buren},
\author[fzk]{A. Weindl},
\author[fzk]{J. Wochele}, 
\author[pol]{J. Zabierowski}

\address[fzk]{Institut f\"ur Kernphysik, Forschungszentrum Karlsruhe, Germany}
\address[uni]{Institut f\"ur Experimentelle Kernphysik, Universit\"at 
Karlsruhe, Germany}
\address[kas]{Institute of Physics and Technology, Almaty, Kazakhstan}
\address[rum]{National Institute of Physics and Nuclear Engineering, 
Bucharest, Romania}
\address[wup]{Fachbereich Physik, Universit\"at Wuppertal, Germany}
\address[pol]{Soltan Institute for Nuclear Studies, Lodz, 
Poland}

\thanks[corr]{corresponding authors, {\it E-mail address:}
haungs@ik.fzk.de; lebedev@satsun.sci.kz}
\thanks[r1]{on leave of absence from Nat.\ Inst.\ of Phys.\ and 
Nucl.\ Eng., Bucharest, Romania}
\thanks[now]{now at: Radboud University Nijmegen, The Netherlands}
\thanks[r4]{on leave of absence from Moscow State University, Russia}

\begin{abstract}
On basis of the theorem of a universal shower development stating
that a hadronically generated extensive air shower is completely 
described by the primary energy, the position of the shower 
maximum and a parameter related to the total muon number, 
the so-called correlation curve method is developed and applied to 
KASCADE data. 
Correlation information of the muon and electron content of showers 
measured by the KASCADE experiment are used for the reconstruction 
of energy and mass of primary cosmic rays. 
Systematic uncertainties of the method and the results are discussed 
in detail. It is shown that by this method general 
tendencies in spectrum and composition indeed can be revealed, 
but the absolute normalization in energy and mass scale requires 
much more detailed simulations. 
\end{abstract}

\end{frontmatter}

\section{Introduction}

Due to the rapidly falling intensity with increasing energy, 
cosmic rays of energies above $10^{15}\,$eV can be studied only 
indirectly by observations of extensive air showers (EAS), 
which are produced by interactions of cosmic 
particles with nuclei of the Earth's atmosphere.  
The observation of a kink in the power law~\cite{kulikov} 
of the size spectrum of EAS and consequently of the all-particle 
energy spectrum at $\sim 3\cdot 10^{15}\,$eV has induced 
considerable interest and  experimental activities. 
Nevertheless, despite of about 50 years of EAS measurements, 
the origin of this so-called 'knee' in the spectrum has not yet 
been convincingly explained~\cite{rpp}. 
In reference~\cite{hoerandel} an overview is given on current models 
trying to explain the origin of the knee.  
Many of these models predict a detailed shape of the primary 
cosmic ray spectrum around the knee with a specific variation of 
elemental composition.
The experimental access to understand the knee requires
accurate measurements of the energy spectra of individual cosmic 
ray elements.

The strategy pursued by the KASCADE collaboration invokes 
an unfolding procedure of the two-dimensional shower size 
spectrum (total electron number vs. EAS muon number) into energy spectra 
of five individual mass groups~\cite{holger}. 
Despite the success of this method for the reconstruction 
of the shape of the spectral forms, a large uncertainty was 
found due to the strong dependence on the hadronic interaction 
models underlying the analyses. 
More general, any interpretation of EAS data obtained from particle 
detectors on ground trying to deduce the energy and composition 
introduces uncertainties of hadronic interactions during the 
shower development.
On the other hand, following the theorem of the shower 
universality~\cite{chou}, each EAS can be characterized by three 
parameters only: the primary energy $E_0$, the depth of the shower 
maximum $X_{\rm max}$, and a parameter describing the muon 
component of the shower, e.g.~total muon number $N_\mu$. 
It is assumed that universal functions exist depending on these three
parameters parameterizing the full shower development, 
i.e.~from these three parameters (corrected for shower-to-shower 
fluctuations) the elemental composition can be extracted and/or constraints 
can be given for hadronic interaction models.
Even more, this theorem professes that the link between primary mass 
and energy to certain shower observables is universal, i.e. independent of 
a specific hadronic interaction model~\cite{giller,nerling}. 

The present work is directed to investigate the possibility and 
accuracy of ground based EAS measurements (in case of KASCADE) 
of extracting elemental energy spectra of individual mass groups by 
assuming the simple assumptions of a universal shower development.  
Full Monte-Carlo shower simulations including the KASCADE detector
response are used for detailed tests of the capabilities of the 
applied method.

\section{Universality of the shower development}

The phrase of an air shower universality states that hadronically
generated showers are in many aspects similar in their 
development through the atmosphere. 
Therefore, a set of a few (e.g.~$E, X_{max}, N_\mu$) 
physical parameters is sufficient to investigate the composition
of the primary cosmic rays and the characteristics of high-energy
hadronic interactions.  
Previous studies have shown, that analytical or semi-empirical 
models for describing the shower development assume or reproduce 
this shower universality. For example the Heitler 
toy-model~\cite{heitler}, which is based on a splitting approximation 
of electromagnetic cascades, relates the shower maximum and the 
number of charged pions to the primary energy 
($X_{\rm max} \propto \log E_0$, $N_{\pi^\pm} \propto N_\mu 
\propto E_0^\beta$).
Investigations with sophisticated Monte Carlo codes for the air
shower simulation on experimentally observable parameters have
demonstrated that characteristics of the electromagnetic component
are universal~\cite{nerling,gora} as well as that the universality 
can be used to normalize the response of a ground detector to the 
muon content of EAS~\cite{schmidt}.  

Recently the basic idea of Heitler was invoked by 
Matthews~\cite{matthews} to reveal salient air shower 
characteristics, where the shower properties are 
related to $E_0 = c_1 (N_e + c_2 N_\mu)$ and 
$X_{\rm max}^p = X_0 + c_3 E_0$ and 
$ln{A} \propto X_{\rm max}^p - X_{\rm max}^A$, 
where $X_0$ is the position of the first
interaction, $X_{\rm max}^p$ the position of the shower maximum 
for primary protons, and $X_{\rm max}^A$ the shower maximum position 
for a particle with primary mass A, respectively. 
Beside the basic parameters describing the EAS
the formulas contain constants $c_i$, the values of which are given by 
particle physics including interaction lengths, elongation rates and 
inelasticity constants. As not all of these constants are known with
high accuracy from accelerator data they also express 
the still existing and presently unavoidable model dependence 
of the interpretation of air-shower data.

To obtain the three basic EAS parameters ($E, X_{max}, N_\mu$) 
sophisticated air shower 
experiments perform hybrid measurements, i.e. measuring
simultaneously the particles on ground and the longitudinal air 
shower development by fluorescence light detection. The energy 
$E_0$ is then deduced either by the total amount of light 
or by the total amount of secondary particles on ground. 
Direct access to $X_{\rm max}$ is provided by fluorescence 
measurements; the muon number has to be obtained by ground or 
underground measurements. 

By classical air shower experiments, i.e. arrays of surface 
particle detectors only, however, $X_{\rm max}$ is not available.  
For such experiments the assumptions of a universal shower 
development have to be introduced to reconstruct the cosmic ray
composition. In most cases this is done by the use of detailed Monte
Carlo simulations. 
However, following Matthews' argumentation, the mass $A$ of the 
primary particle can be estimated less model dependent
by ground observables if the difference in $X_{\rm max}$ of a 
certain primary mass A to $X_{\rm max}^p$ is known.
The assumption of shower universality is suggesting, that the EAS 
development is the same for all showers after the shower maximum. 
Hence, the difference of electron number and muon number at 
observation level should be a direct measure of the differences in 
depth of the shower maximum and consequently of the primary mass.      
This is the principal idea of the present studies.
 
In this approach we will analyze KASCADE data with the 
so-called correlation curve 
approach where ground observables are used. To reduce effects of 
shower-to-shower fluctuations the correlation between the measured 
observables are also taken into account instead of using  
above mentioned formulas only. For that, the original 
correlation curve method~\cite{boos,lebedev} 
was modified and as relevant parameters 
the correlation of the electron, $N_e$, and muon, $N_\mu$, 
number is considered. The modification concerns the exchange of the 
lateral slope of the particle distribution ('age') with the
number of shower muons, as the age measured close to sea-level was 
found to loose most of its information on the longitudinal
development. The results will be compared for different ranges 
in zenith angle, for different hadronic interaction models 
underlying the analysis, and with the results of the KASCADE 
unfolding procedures~\cite{holger}.

\section{KASCADE experiment}

The KASCADE experiment located at the site of the northern 
campus of the Karlsruhe Institute of Technology KIT, Germany, 
measures various observables of extensive air 
showers with primary energies between $3\cdot 10^{14}$ and 
$1\cdot 10^{17}$ eV. 
It consists of three major detector components: 
the field array, the muon tracking detector 
and the central detector complex~\cite{kascade}.
For this analysis only the field array is of relevance, which 
extends over an area of $40.000\,$m$^2$ and consists of 252 
detector stations for detection of the electron component, where 
192 stations employ in addition shielded scintillators for  
muon measurements. 
The data set used in the present analysis is similar to that one 
used for the unfolding analyses described in~\cite{holger} 
and~\cite{isv06}. Details of the reconstruction 
of the shower axis direction, the total electron number and the 
truncated muon number (muon content in $40-200\,$m core distance, 
used for the present analysis) are described in~\cite{kascade} 
and~\cite{lat}.

\section{Simulations}

The simulations of the EAS development have been performed by 
using the QGSJet01 high-energy hadronic interaction model~\cite{6} 
in the frame of the CORSIKA program version 6.156~\cite{7}. 
For the low energy interactions ($E_{cm}<200\,$GeV) the 
FLUKA~\cite{15} (version 2002.4) code, and for 
treating the electromagnetic part the EGS4~\cite{egs} program 
package was used.
About 2 millions of EAS in the energy interval of $10^{14}\,$eV 
to $10^{18}\,$eV for each of 5 primaries (p, He, 
C, Si, Fe) have been simulated. The energy distribution follows a 
power law with slope index of -2. The zenith angles are distributed 
in the range $[0-42^\circ]$.   
In order to take into account the installation response a detailed 
GEANT~\cite{8} simulation of the KASCADE detectors and the 
reconstruction by the standard KASCADE reconstruction software 
was used. 
With less statistics the simulations were repeated with the SIBYLL 
(version 2.1)~\cite{sibyll} code as high energy interaction model 
instead of QGSJet.

\section{Correlation curve method and simulation studies}

The sensitivity of most of EAS observables measurable at ground level 
to the mass of the primary cosmic ray particles is rather weak due to huge 
fluctuations during the shower development in the atmosphere, which can be 
even larger than the difference of the mean values of primary protons and 
iron nuclei. 
The main contribution to these fluctuations of $N_e$ and $N_\mu$ at a 
given observation level is connected with parameters of the first or 
first few interactions in the atmosphere. 
For example, if a 100 PeV cosmic iron comes into collision with nitrogen 
of the Earth's atmosphere the multiplicity of the interaction can be only
a few secondary particles (peripheral or diffractive interaction) or up to 
hundreds of secondaries (central collision). 
Each of these secondary particles interact further with 
nuclei of the atmosphere and therefore can produce secondary shower
particles in a wide interval of multiplicity.
After a few generations of interactions the further development of the EAS 
gets more similar due to a averaging effect and the shower behaves 
significantly less fluctuating. 

In addition, the reconstruction of EAS gets more difficult by combining  
different zenith angles, because EAS impinging with 
large zenith angles cross a thicker layer of the atmosphere and
consequently vary in $N_\mu$ and $N_e$. 
Hence, a correction of $N_e$ and $N_\mu$ is necessary, which leads to 
additional uncertainties.

A possible approach to overcome this slant depth effects 
can be found by correlation analyses of different observables. 
In this work, a correlation analysis 
of the number of muons, $N_\mu$, and electrons, $N_e$ in the EAS is 
considered.
By assuming the shower development universality, in contrast to the first 
high-energetic hadronic interactions the electromagnetic development 
is similar for all showers: If showers are generated with same incident 
parameters (energy, mass), they have a similar development - after the 
shower maximum. Therefore, the correlation of the electron and 
muon number at observation level hints to the position of the shower 
maximum, i.e. in particular to the behavior of the first interactions.

\subsection{The correlation curve method}

A correlation curve is defined as a polynomial function in 
a two-observable plane fitted to full simulated showers of 
same primary characteristics.  
\begin{figure}[b]
\begin{center}
 \includegraphics[width=0.495\linewidth]{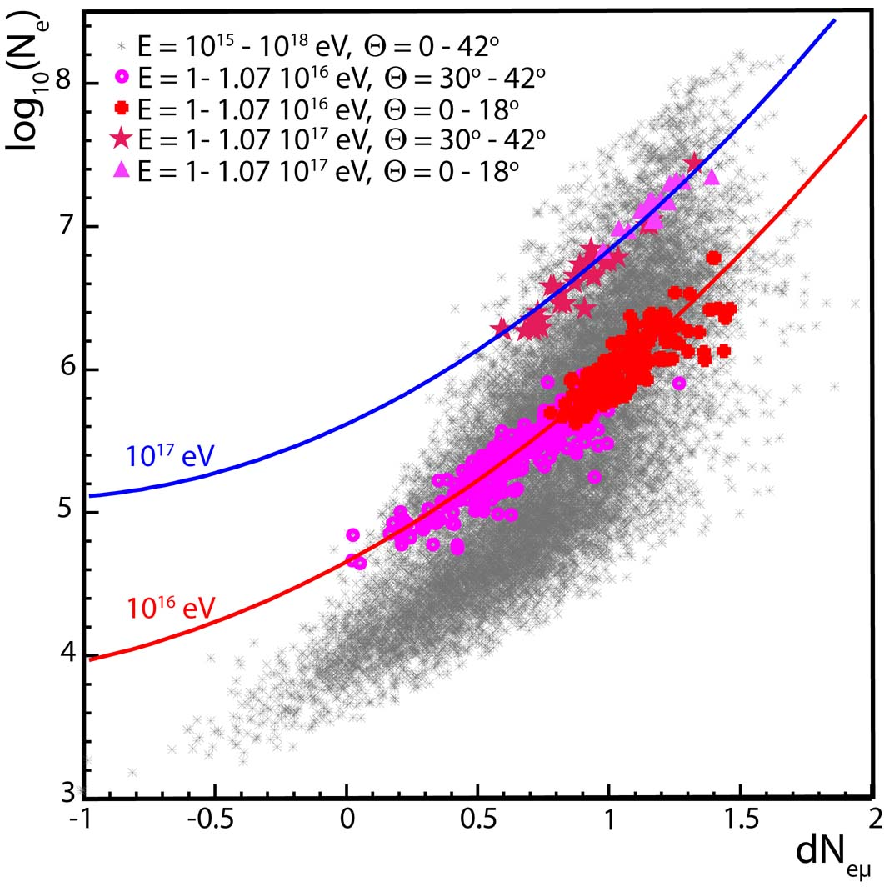}
 \includegraphics[width=0.495\linewidth]{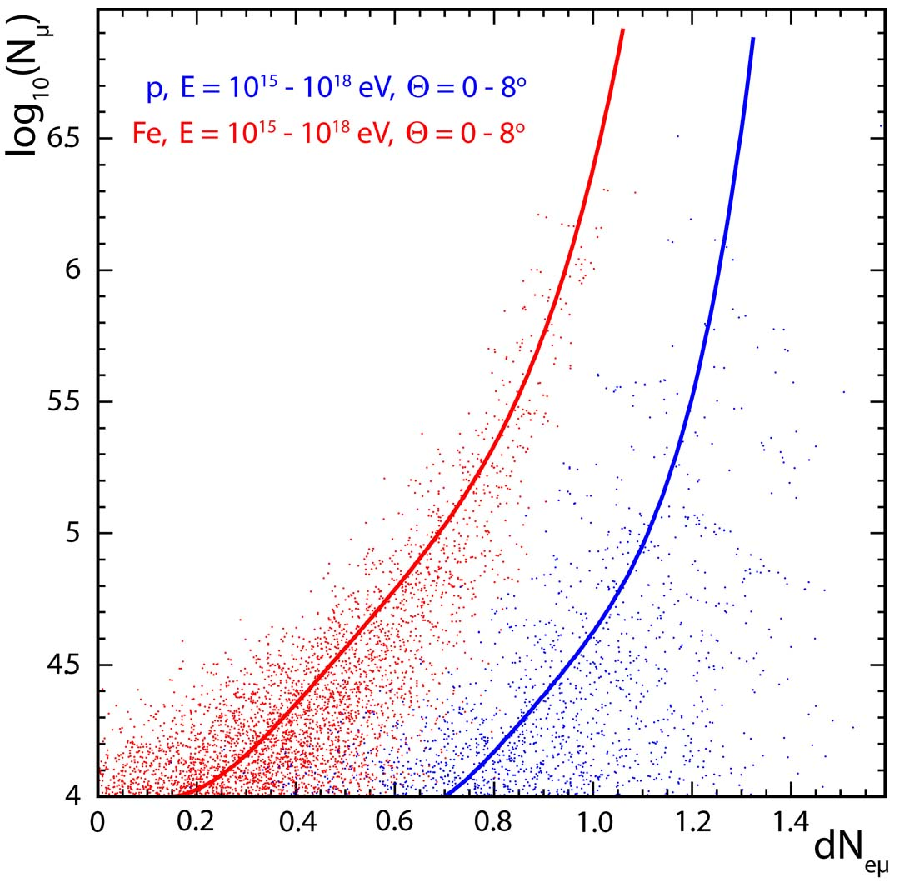}
   \caption{Left: $log_{10}(N_e)$ vs. $dN_{e\mu}$-distribution for proton 
initiated EAS, simulated by QGSJet including the KASCADE detector response, 
and example correlation curves as used for the energy reconstruction.
Right: $log_{10}(N_\mu)$ vs. $dN_{e\mu}$-dependences for iron and 
proton initiated showers and correlation curves as used for the 
mass reconstruction.}
	\label{correl}
\centering
\end{center}
\end{figure}

For the energy reconstruction this methodical approach uses 
$log_{10}(N_e)$  vs. $dN_{e\mu}$ correlation curves, where  
$dN_{e\mu}=log_{10}(N_e)-log_{10}(N_\mu)$. 
Fig.~\ref{correl} (left) shows, that all showers of the same mass and 
energy in a wide interval of zenith angles ($0-42^\circ$)
concentrate around specific $log_{10}(N_e)$ vs. $dN_{e\mu}$ 
correlation curves.  
In this presentation, the large fluctuations of $log_{10}(N_e)$ are 
reduced to smaller fluctuations around the correlation curves.
By this, the use of such correlation curves allows to suppress the  
influence of the fluctuations (connected with parameters of the 
first interactions) on the energy reconstruction.
A further positive aspect of this approach lies in the small 
dependence on the zenith angle (at least for $<42^\circ$) of the shower. 
Early developing EAS with small zenith angles have 
at observation level values of $N_e$ and $N_\mu$ which are close to $N_e$ 
and $N_\mu$ values of later developing showers with larger zenith angles.  
Hence, showers with large zenith angles help to obtain a more extended 
correlation curve with therefore a smaller uncertainty which leads to a 
more accurate determination of the primary energy. 
Fig.~\ref{correl} (left) illustrates this by showing showers 
with different zenith angles. 

Whereas $log_{10}(N_e)$ vs. $dN_{e\mu}$ correlation curves are used 
for energy reconstruction, for mass reconstruction 
$log_{10}(N_\mu)$ vs. $dN_{e\mu}$ correlation curves are proposed. 
Most of the showers of the same mass in a wide interval of primary energy 
($10^{15}-10^{18}\,$eV, at least) are placed 
around such curves (see Fig.~\ref{correl}, right).
The $log_{10}(N_\mu)$ vs. $dN_{e\mu}$ correlation 
curves differ significantly for proton and iron showers.  
This allows to separate proton and iron showers (at least) with high 
efficiency in almost the full energy interval 
(at least, from $5\cdot 10^{15}$ to $10^{18}$ eV).
By simulations it was found that the largest source for uncertainties 
in the mass reconstruction at low energies ($log(E/GeV)<6.5$) is due 
to uncertainties of the $N_{\mu}$ reconstruction. 
When a shower covers a small 
area only and the number of hit muon detectors is small, the statistical  
uncertainties of $N_\mu$ reconstruction increase and therefore the
uncertainties of the mass reconstruction increase.

\subsection{Procedure of mass and energy reconstruction}

The final procedure of the energy and mass reconstruction in the present
studies is realized in the following way: 
\begin{figure}[b]
\begin{center}
 \includegraphics[width=0.495\linewidth]{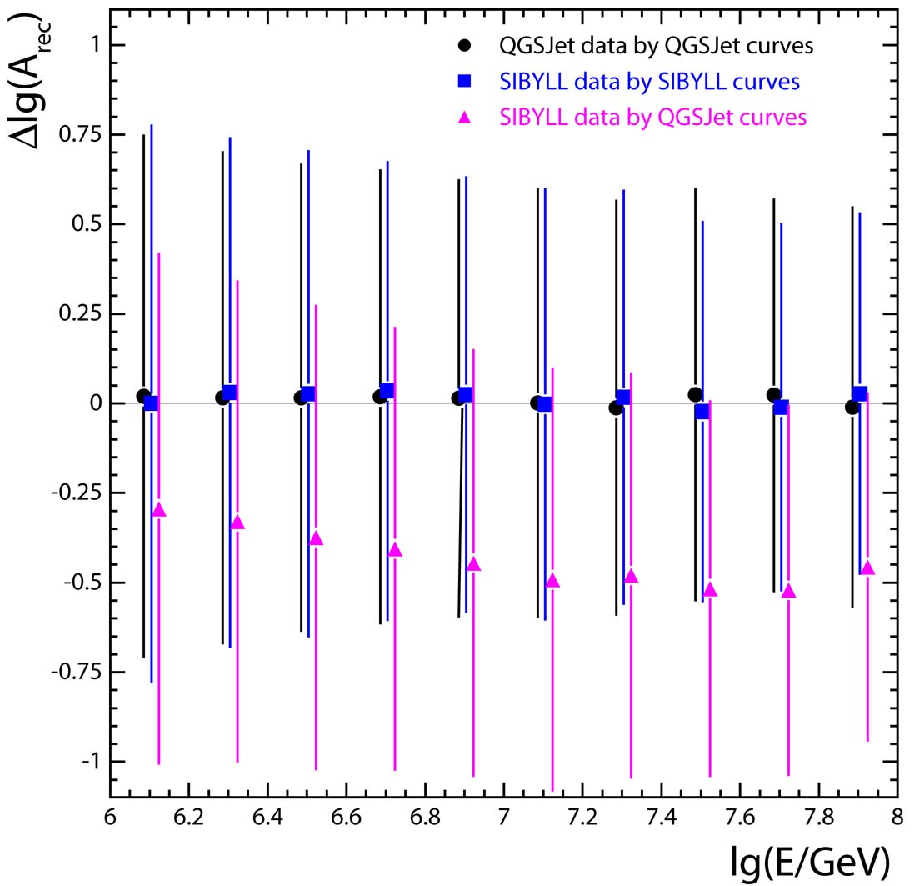}
 \includegraphics[width=0.495\linewidth]{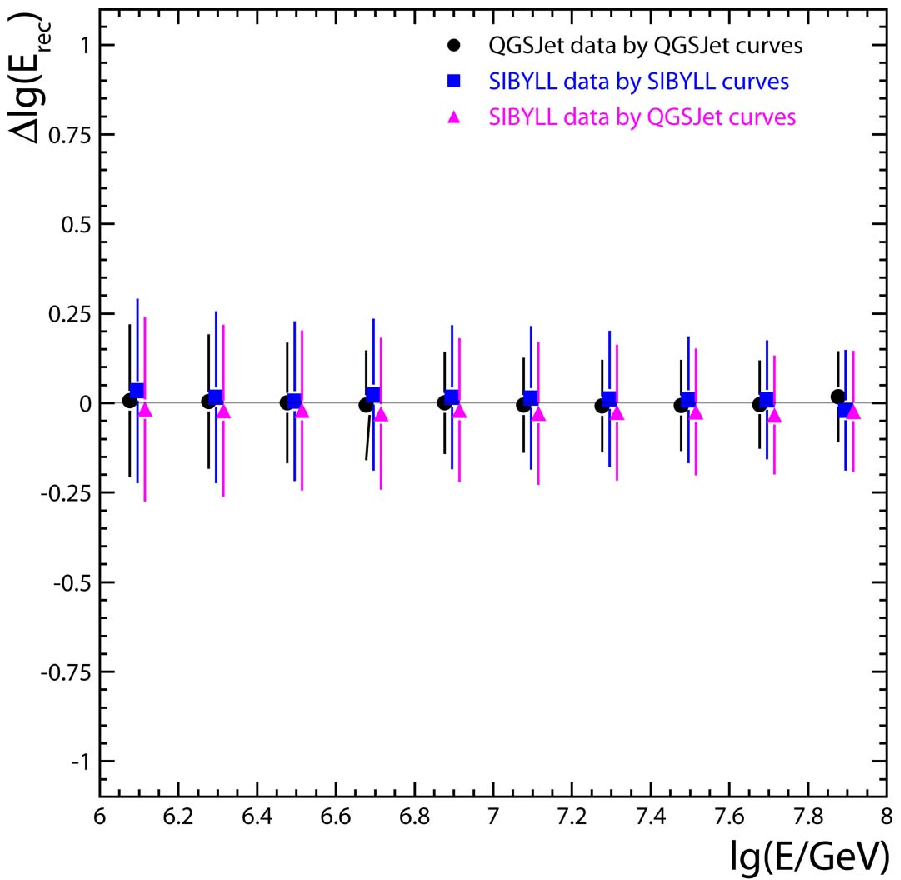}
   \caption{Left: Accuracy in mean mass reconstruction shown for QGSJet
   simulations by applying correlation curves obtained by QGSJet
   simulations, for SIBYLL simulations by applying correlation
   curves obtained by SIBYLL simulations, and for SIBYLL simulations
   by using QGSJet for determining the correlation curves. Used are the full
   Monte Carlo sample of 5 primary mass groups with equal abundances. 
	For a better visibility some markers are displaced.
	Right: Accuracy in energy reconstruction. Same notation as in
	left figure.}
	\label{accur}
\end{center}
\end{figure}

\begin{enumerate}
\item For each of the simulated primaries the $dN_{e\mu}$ vs. 
$N_\mu$-distribution in narrow bins of zenith angles 
is fit by a polynomial function taking $\theta$ and  $A$ as 
free parameters. This yields a function $dN(N_\mu, A, \theta)$. 

\item The mass of the showers is determined from the function by 
varying $A$ and minimizing the difference between the measured 
$dN_{e\mu}$ and the function $dN(N_\mu, A, \theta)$ at fixed 
(measured) $\theta$ and $N_{\mu}$.

\item  Using simulations for different fixed energies and masses  
the $log_{10}(N_e)$ vs. $dN_{e\mu}$ - distributions are fit with a 
polynomial function taking $E_0$ and  $A$ as free parameters. 
This yields a function $log_{10}(N_e(dN_{e\mu}, E_0, A))$. 

\item The primary energy is determined from this function by varying 
$E_{0}$ and minimizing the difference between the measured 
$log_{10}(N_e)$ and the function 
$log_{10}(N_e(dN_{e\mu}, E_0, A))$ at the fixed (measured) 
$dN_{e\mu}$ and fixed (reconstructed) $A$.
\end{enumerate}

Figure~\ref{accur} shows the obtained accuracies for the mass and energy 
reconstruction by this procedure. 
While the energy reconstruction shows 
only a small dependence on the interaction model and a smaller width 
of the distribution, the dependence on the models and the width 
is larger in case of the mass reconstruction.
In particular, if one fits the correlation curves based on one model and 
applies these curves to a test data set generated by the other model, 
a clear systematic shift is observed. 
Inside one model the reconstruction is accurate, meaning the shower 
universality is preserved, even after detailed simulations including 
the full chain of Monte Carlo shower simulation, detector simulation, 
and the reconstruction. 
This behavior is not surprising as the constants in, for example, 
Matthews's interpretation of Heitler's approximation depend on the 
model of the hadronic interactions. 
In addition, this is the basis of differences between QGSJet and 
SIBYLL.

In the simulation program CORSIKA 
different hadronic interaction models are used for high energy 
($E_{lab}>200$ $GeV$) and low energy ($E_{lab}\le 200$ $GeV$) interactions. 
Hence, the high energy models control especially the first few interactions 
of an air shower, whereas most of the (electromagnetic) particles detected 
at ground level are generated by the low energy interaction models. 
Two important parameters characterizing the high-energy interaction models 
are the inelastic cross section and the elasticity of the interaction.  
A lower cross section implies a longer mean free path for the hadrons in the 
atmosphere and thus a reduction of number of interactions. Showers develop 
more slowly and the mean $N_{\mu}/N_{e}$ ratio at ground level decreases. 
A lower elasticity implies that less energy is transfered to the leading 
particle. It enhances the possibility to produce secondary particles as
more energy is available for multi-particle production. Showers develop 
more quickly and the mean $N_{\mu}/N_{e}$ ratio at ground level increases.

SIBYLL has a larger (compared to QGSJet) inelastic cross 
section and a larger elasticity. This means that we are faced with two 
competing processes influencing the number of particles observed and 
the position of the shower maximum. 
Whereas, the depth of the shower maximum, predicted by SIBYLL and QGSJet
are close to each other, SIBYLL 2.1 simulations predict about 8\% 
smaller muon number and 10-20\% larger electron number than 
QGSJet 01~\cite{milke}. 
These differences in the predictions of muon and electron numbers 
at the observation level lead to the significant effect for the mass 
reconstruction. Consequently, investigations with further 
high-energy hadronic interaction models 
(e.g. the newly developed EPOS model~\cite{epos}), 
with again different predictions on the electron-muon number ratio, 
would lead to varying mass reconstruction and therefore varying 
results in the interpretation of measured data in means of composition.
The influence of using different low energy interaction models 
(FLUKA and GHEISHA) was found to be negligible for the discussed 
observables~\cite{isv06}.

\subsection{Energy spectra of individual mass groups}
       
\begin{figure}[ht]
\begin{center}
\includegraphics[width=0.495\linewidth]{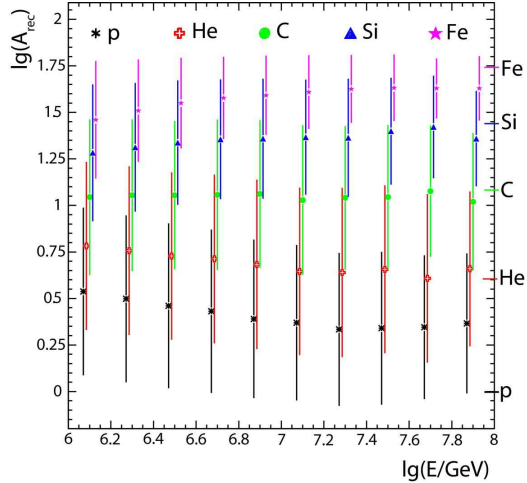}
\caption{Resolution in mass reconstruction (QGSJet based). 
Equally weighted number of primaries are generated 
and the resulting mean mass lg(A) and the width of the distributions 
are shown in dependence of the primary energy.}
	\label{resol}
\end{center}
\end{figure}

A crucial element in reconstructing energy spectra of individual masses 
(mass groups) is the intrinsic mass resolution of the used observables and 
of the method applied.
Whereas the energy reconstruction is based mainly on the total 
number of particles, the mass reconstruction is based on the correlation 
of the electron and muon component leading to much larger uncertainties. 
Figure~\ref{resol} depicts the mass resolution in case of QGSJet and shows, 
that with the simple ansatz used, the medium masses are not satisfactorily 
resolved. 
We remark here, that this feature was the reason to go a step 
forward to a more sophisticated investigation of the  event-by-event 
correlations, resulting in the analysis described in~\cite{holger}. 
But it is also seen, that by all means, a separation in two mass groups 
light and heavy is possible. 
Hence in the further application, we divide the data set in two parts: 
in a sample of light induced showers ('L') with a reconstructed 
mass of $A < 12$, and a sample of heavy induced showers ('H') 
with $A\ge12$, respectively. 
This mass cut is chosen in a way that the CNO group is separated to 
$\approx 50$\% in the light and the heavy group, respectively.  

\begin{figure}[ht]
\begin{center}
\includegraphics[width=0.495\linewidth]{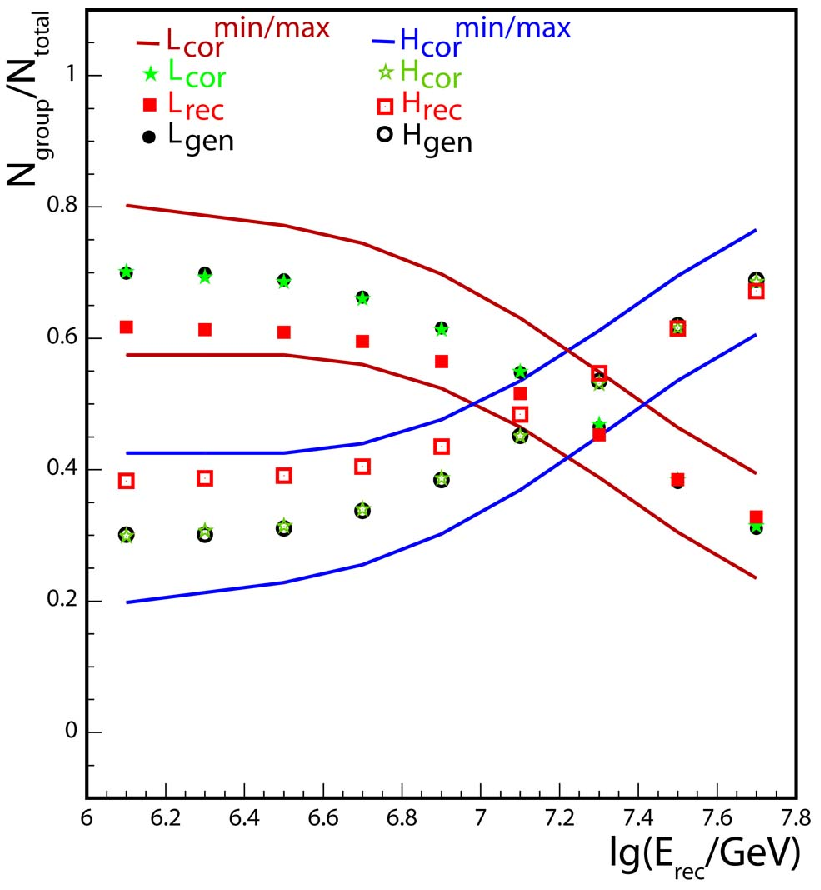}
\includegraphics[width=0.495\linewidth]{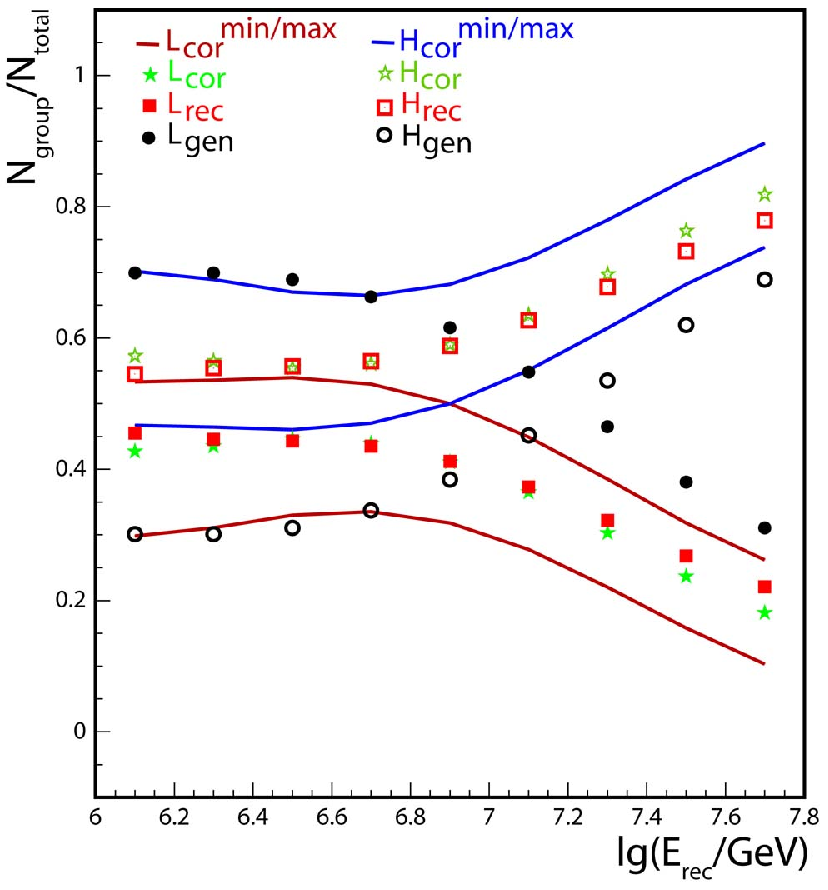}
\caption{Comparison of simulated (index {\it gen}) and 
reconstructed elemental mass spectra (light and heavy 
in terms of relative abundances) before (index {\it rec}) 
and after (index {\it cor}) correction for the mis-classification. 
The lines limits the maximal possible uncertainty due to the composition
within the mass groups. Statistical uncertainties are not shown.
Left: QGSJet based (generation and reconstruction). Right: SIBYLL based 
reconstruction, but QGSJet based generation of the test sample.}
	\label{mctest}
\end{center}
\end{figure}

Before we apply the correlation curve method to the measured data, 
the reconstruction procedure is tested with an artificial test sample.
For that we use as input in gross features the unfolding result 
of~\cite{holger} (based on QGSJet), i.e. the spectra and relative 
abundances of five primary mass groups, for the present studies grouped
in ratios of light and heavy primaries, respectively,  
to the total amount of events. This input composition is displayed in 
Fig.~\ref{mctest} with index {\it gen}. 
The application of the reconstruction steps to this test sample results 
in the reconstructed relative numbers of heavy and light as displayed in 
Figure~\ref{mctest} (index {\it rec}).
For obtaining a reliable result still missing is the deconvolution of 
systematic uncertainties in mass reconstruction and energy estimation 
leading to a correction of the reconstructed individual mass spectra. 
The mis-classification probabilities are obtained from simulations with 
the assumption of equal number of primaries in four generated mass groups 
As a first step a test sample is considered with equal abundances of 
Fe and Si forming the heavy group and equal abundances of p and He 
forming the light group. In order to check the accuracy of the method, 
we repeated these tests comparing output data with different input compositions 
and by that we calculated the maximal possible uncertainty due to the 
composition within the individual mass groups. 

Thus, the final results are the spectra corrected for the energy-dependent 
misclassification probabilities (index {\it cor}).  
The upper and lower lines in Fig.~\ref{mctest} limit the maximal uncertainty 
of this correction due to the unknown, but assumed composition inside the light 
and heavy groups (ratio of protons to Helium and Si to Fe, respectively for
the calculation of the mis-classification probabilities).
Where in the left part of the figure everything is based on the QGSJet model,
for the right part the result is shown if the input spectra are generated with 
QGSJet and the whole reconstruction is based on values obtained by using the 
SIBYLL model. These investigations show that the shape of the distribution is 
preserved but SIBYLL reconstructs a heavier composition. 
Comparing the test results for QGSJet and SIBYLL, it is obvious that the 
largest difference is due to the hadronic interaction model.  

In summary, the correlation curve method is able to 
confirm the shower universality in basic considerations.
This finding is due to the fact, 
that by investigating the total number of secondary particles together with 
the difference in electron and muon number at ground level it is possible
to reconstruct the primary energy and to classify in two mass groups, at least.
In addition, the method avoids to a large amount the problem of correction 
of $N_e$ and $N_\mu$ for EAS with 
different zenith angles (and uncertainties connected with the correction). 
The largest deficiency of the described reconstruction procedure is 
due to the large shower-to-shower fluctuations, anyhow not considered in the 
assumptions of simple Heitler-based models.

\section{Application to KASCADE data}

The correlation curve method as described in the last chapter is 
applied to the KASCADE data set, where  a `good-run'-selection 
of KASCADE is used. 
In these runs all clusters and all detectors are present and working and are 
well calibrated. In total the used sample sums up to about 993 days 
measuring time. This is the same sample as used in ref.~\cite{holger}, the
KASCADE unfolding analysis of the two-dimensional shower size spectrum. 

\begin{figure}[ht]
\begin{center}
\includegraphics[width=0.495\linewidth]{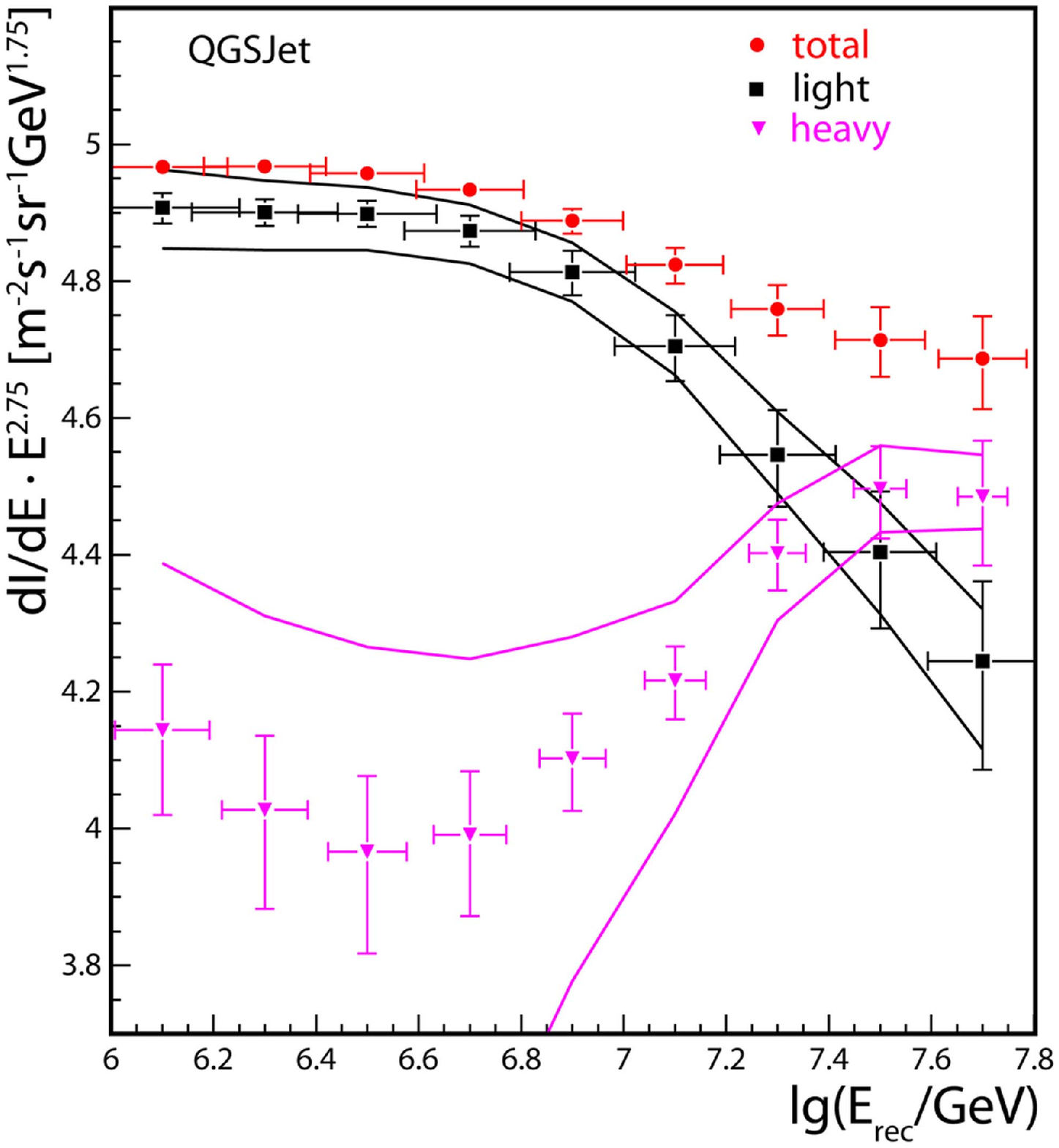}
\includegraphics[width=0.495\linewidth]{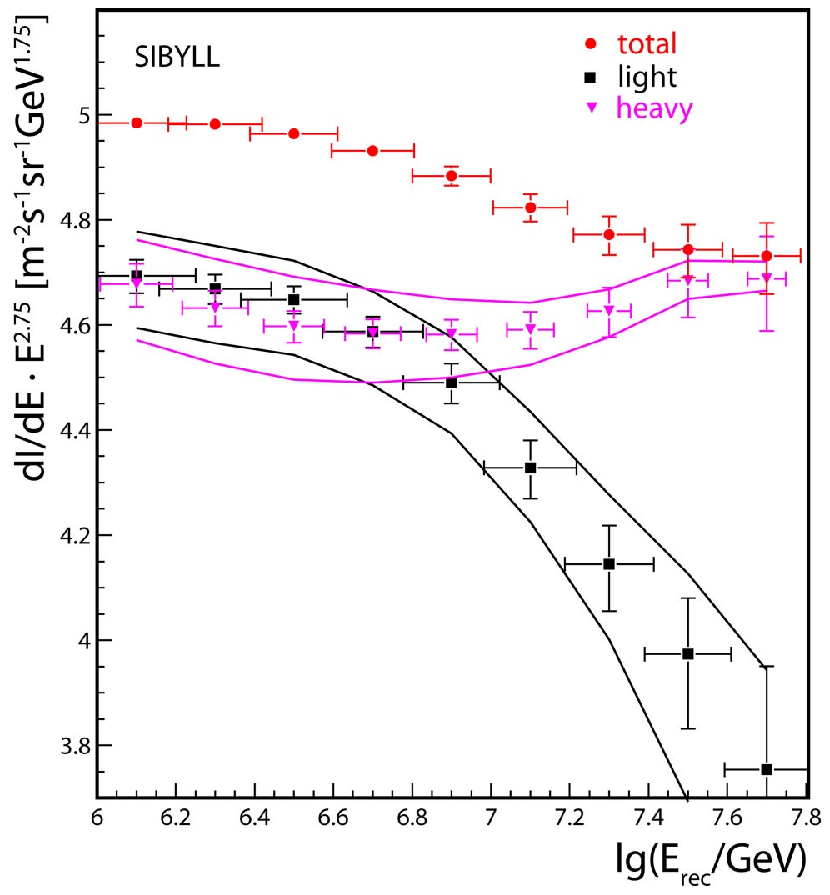}
   \caption{KASCADE energy spectra reconstructed for the zenith angular 
   interval of $\theta =0-18^\circ$, calculated by applying the correlation 
   curve method on the basis of the QGSJet-FLUKA model (left) and 
   based on the SIBYLL-FLUKA (right) model. For the description of the error
   bars see text. }
	\label{result}
\end{center}
\end{figure}

In Figure~\ref{result} the reconstructed all-particle spectrum is shown 
together with the spectra of light and heavy induced showers for the 
zenith angular range of $0-18^\circ$.
The left panel of the figure shows the result by using QGSJet based 
simulations, the right panel for the SIBYLL model.
The horizontal error bars denote the uncertainty in energy reconstruction.
The vertical error bars combine the statistical uncertainty of both, data and 
simulations with the uncertainty due to the misclassification probabilities.
The lower and upper limits shown in the plots describe the maximum 
uncertainty due to the assumption of a composition in the calculation of the 
mis-classification probabilities.
 
Knee like features are clearly visible in the all-particle spectrum of 
both results, QGSJet and SIBYLL based, as well as in
the spectra of light primaries. This
demonstrates that the elemental composition of
cosmic rays is dominated by light components below
the knee and dominated by a more heavy component above the knee feature. 
Thus, the knee feature originates from a decreasing flux of the light
primary particles. 
This observation corroborates results of the analysis of the unfolding 
procedures described in~\cite{holger} and of muon density measurements
at KASCADE~\cite{muon}, which were performed
independently of this analysis, as well as results from the 
EAS-TOP~\cite{eastop} experiment.

Comparing QGSJet based with SIBYLL based results, the all particle 
spectrum as well as the characteristics of the individual mass group spectra 
\begin{figure}[ht]
\begin{center}
 \includegraphics[width=0.495\linewidth]{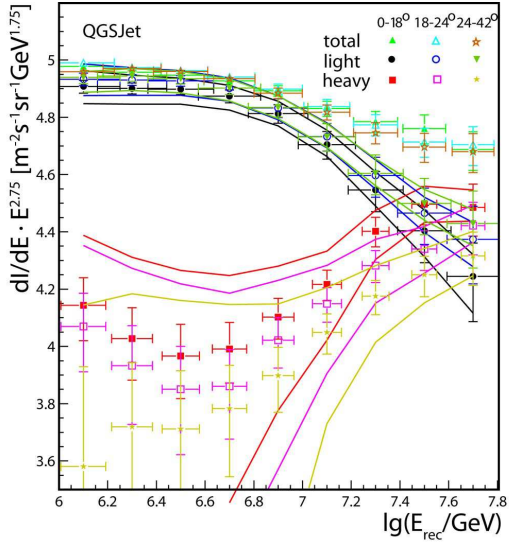}
   \caption{Comparisons of the QGSJet based results for different
   angular ranges.}
   \label{angle}
\end{center}
\end{figure}
(distinct knee feature for the light components, but not for the heavy ones) 
are in remarkable agreement, but there are large differences in the 
relative abundances of the different primaries. 
This confirms that the assumption of a universal shower 
development is valid, in particular for the evolution of the shower 
in the atmosphere. This is not true for the absolute number of particles 
on ground, where large differences in the model predictions occur.
It is obvious that the obtained individual energy spectra 
(especially for the relative abundances of the different mass groups)
depend also on the reconstruction method and/or the low energy 
interaction model, but the main differences stem from varying the 
high-energy hadronic interaction model.  
\begin{figure}[th]
\begin{center}
 \includegraphics[width=0.85\linewidth]{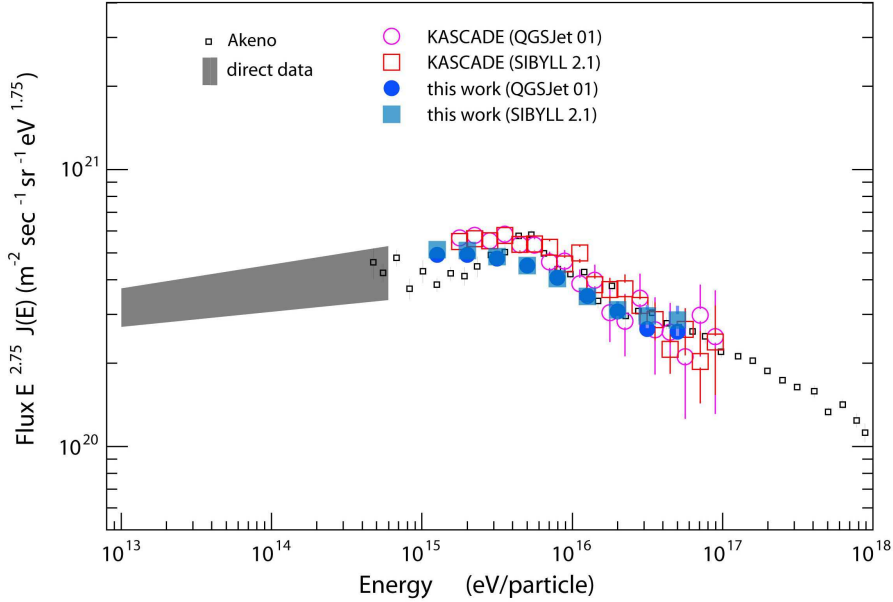}
   \caption{The all-particle spectra obtained on basis of two
   different hadronic interaction models. Only statistical error bars
   are shown.  
   Results of the KASCADE unfolding analysis~\cite{holger} 
   and the AKENO experiment~\cite{akeno} are also shown, as well as 
   a compilation of direct measurements~\cite{watson}.}
   \label{comp}
\end{center}
\end{figure}

To deepen the understanding and as cross-check of the source of the 
differences in the two results, the correlation curve method is 
applied to data in different zenith angular bins. 
Fig.~\ref{angle} displays the resulting energy spectra for different ranges
in zenith angle (in case of QGSJet). 
The all-particle spectrum as well as the relative abundances of the different 
mass groups are in good agreement, confirming differences in the different
hadronic interaction models as main source of diverging results.

Finally, in Figure~\ref{comp} the resulting all-particle spectra are
compared with the results based on the unfolding
procedure~\cite{holger} with the same high energy interaction models
and on the same data set ($0-18^\circ$). Differences are due to the
reconstruction method and the use of a different low energy interaction
model (FLUKA here instead of GHEISHA), where the influence on the low
energy interaction model was found to be of minor importance~\cite{isv06}.
In addition, a compilation of direct measurements~\cite{watson} and
the results from the AKENO~\cite{akeno} experiment are included as well. 
The present result on the all-particle spectrum is essentially independent 
of the interaction model used and in good agreement
with results from other experiments. 

\section{Discussion and conclusions}

Based on the assumption of a universality of the shower development
in air, observables were extracted measurable by a ground based
particle air shower experiment.
The theorem of shower development universality asserts that the shower
can be fully described if the energy, the position of the shower
maximum and a normalized characteristics of the muon content of the
shower are known. In particular, by having these values, the primary
mass and the mechanisms of the first or few first highest energy 
interactions can be inferred. 

In this work, investigating KASCADE data and therefore 
relying on ground observables, only, we defined
the logarithmic difference of ground measured electron and muon
numbers as the parameter connected with the position of the shower
maximum. This connection is grounded on studies of 
Matthews~\cite{matthews} revealing salient air shower 
characteristics by simple formulas based on the assumptions of a
universal shower development.  
As method of reconstructing the primary mass and energy 
of the showers the correlation curve method is applied taking into 
account not only the total electron and muon number of the shower, but 
also the correlations between the used observables. 
In addition, using invariant correlation curves of 
reconstructed $N_e$ and $N_\mu$ allows significantly to decrease the 
influence of shower-to-shower fluctuations on the 
reconstruction of energy and mass of the primary cosmic rays.
Uncertainties of the energy reconstruction are almost independent 
of the slope of initial energy spectrum and initial mass composition. 
Furthermore, in this approach the energy 
reconstruction is nearly independent of the zenith angle 
(at least for $\Theta < 42^\circ$). 

By detailed simulation studies the energy and mass reconstruction
accuracies could be estimated, demonstrating that indeed the connection 
between the ratio of electron to muon number to the position of the 
shower maximum is given, and therefore a sensitivity to the primary 
mass is maintained in the ground observables.
I.e., a universality in the shower development is grossly given.
This concerns in particular the ratio of muons to electrons, but is 
less valid for the absolute numbers of particles on ground and 
neglects the effect on shower-to-shower fluctuations.
The description of an universal behavior of the shower development is given 
probably for all the widely used simulation codes and interaction models, 
even if in the present studies this was tested in detail only for the 
models QGSJet and SIBYLL. 

Therefore, to quantify the energy spectrum and mass composition,
hadronic interaction models still are needed. In particular,
the normalization to the absolute value of the energy and to the 
mass have to make use of models.  
In terms of the revised Heitler model~\cite{matthews} it means, that
the constants in the 
given formulas have to be defined, which need input from particle physics.
The relative behavior of the measured showers (shape of the spectrum and
composition changes), however, can be revealed by assuming basic shower 
development considerations, i.e. applying the shower universality.
However, shower-to-shower fluctuations are annoying for a
simple application of the universality theorem. Especially, following
the shower to sea-level, the assumptions are too simple and 
underestimate the fluctuations (which is possibly also true for full 
Monte Carlo simulations).

By applying the correlation curve method to fully simulated KASCADE
EAS, it was found that a proper mass discrimination is possible only
in two mass groups - light and heavy induced showers.
Again the largest uncertainty is in mass estimation and by choosing
different hadronic interaction models, in this case by comparing
QGSJet and SIBYLL. As guessed by the universality theorem, energy
spectra and shape of the relative composition are revealed to be quite 
stable, but the absolute normalization to the mass scale is different 
leading to a much larger (heavier) mean mass in case of SIBYLL. 
Modified or newly developed hadronic interaction models with changes 
of the interaction mechanisms probably would take us to other, 
changed abundances. 
  
Finally the reconstruction procedure was applied to KASCADE data,
resulting in QGSJet and SIBYLL based all-particle spectra as well as
energy spectra of the light and heavy groups. 
The results are in good agreement with other experiments, in
particular with results of the much more sophisticated ansatz of
unfolding the two dimensional size spectrum measured by KASCADE. 

As basic conclusion by this investigation we can confirm that 
the composition gets heavier and heavier crossing the knee in the 
all-particle spectrum due to the fact that the showers become muon-richer. 
This was found in many experiments 
by using detailed simulations~\cite{holger} 
or simple models or assumptions on the universal shower 
development~\cite{muon}, only. 
The present study could show that very strong changes in the description of
hadronic interactions have to occur and completely overruling the shower 
development universality if one would explain the knee without changing 
composition. 
As an example in this direction, in~\cite{alvarez} the effect of percolation, 
i.e. an increase of the energy part into one leading pion, is discussed. 
The mechanism would change the 
inelasticity of the interactions, and therefore the basic correlations of 
shower observables. But the effect, if true,  can be expected for energies 
above 100 PeV, only and is anyhow not strong enough to explain the 
very distinct knee feature.

In summary, assuming the shower development universality to be true in 
its basic features it is clear, that there is a knee in the all-particle 
spectrum at a few PeV, and that this knee is due to a kink in the spectrum 
of light primary particles.
  
\begin{ack}
This work is supported by the Deutsche Forschungsgemeinschaft
(DFG grants GZ436KAS17/1/04 and 436KAS17/1/07) and by the 
Kasach Academy of Science, what is gratefully acknowledged 
by I.L. and A.H..
\end{ack}


\begin{thebibliography}{99}                                                     

\bibitem{kulikov} Kulikov G.V., Khristiansen, Soviet Physics JETP 35 (1959) 441.

\bibitem{rpp} Haungs A., Rebel H., Roth M., {\it Rep. Prog. Phys.} \textbf{66} (2003) 1145.

\bibitem{hoerandel} H\"orandel J.R, {\it Astropart. Phys.} \textbf{21} (2004) 241.

\bibitem{holger} Antoni T. et al. - KASCADE-Collaboration, 
		{\it Astropart. Phys.}  \textbf{24} (2005) 1.

\bibitem{chou} Chou A.S. et al., 2005, 
		Proc.29$^{th}$ ICRC, Pune, \textbf{7} (2005) 319.

\bibitem{giller} Giller M. et al., 2004,  
	{\it J. Phys. G: Nucl. Part. Phys.} \textbf{30} 97. 

\bibitem{nerling} Nerling F. et al., 
		{\it Astropart. Phys.}  \textbf{24} (2006) 421.

\bibitem{heitler} Heitler W., {\it Quantum Theory of Radiation} 
		Oxford University Press (1944).

\bibitem{gora} Gora D. et al., 
		{\it Astropart. Phys.}  \textbf{24} (2006) 484.

\bibitem{schmidt} Schmidt F. et al., 2007, 
		Proc.30$^{th}$ ICRC, Merida (2007), in press.

\bibitem{matthews} Matthews J., {\it Astropart. Phys.} \textbf{22} (2005) 387.

\bibitem{boos} Boos E.G. et al., 2001, Proc.27$^{th}$ ICRC, Hamburg, p.269

\bibitem{lebedev} Boos E.G. et al., 
		{\it Kaz.J. Izv.AS RK ser.fiz.-mat.}  \textbf{N 2} (2002) 68.

\bibitem{kascade} Antoni T. et al. - KASCADE-Collaboration, 
		{\it Nucl. Inst. Meth. Phys. Res. A} \textbf{513} (2003) 490.

\bibitem{isv06} Ulrich H. et al. - KASCADE-Grande-Collaboration, 
		{\it Nucl. Phys. B (Proc. Suppl.}) (2007), in press.

\bibitem{lat} Antoni T. et al. - KASCADE-Collaboration, 
		{\it Astropart. Phys.} \textbf{14} (2001) 245. 

\bibitem{13} Haungs A. et al. - KASCADE-Grande collaboration, 2003, 
		Proc.28$^{th}$ ICRC, Tsukuba, HE1.5, p.985

\bibitem{6} Kalmykov N.N., Ostapchenko S.S., 
		{\it Sov.J.Yad.Fiz.} \textbf{56} (1993) 105.

\bibitem{7} Heck D. et al., FZKA-Report 6019, Forschungszentrum Karlsruhe, 1998

\bibitem{15} A. Fasso et al., Proc. Monte Carlo 2000 Conf., Lisbon, Oct. 23-26, 
		2000, A. Kling et al. eds., Springer (Berlin) 955 (2001).

\bibitem{egs} Nelson W.R., Hirayama H. and Rogers D.W.O., {\it Report} 
		\textbf{SLAC 265}, Stanford Linear Accelerator Center (1985).

\bibitem{8} CERN, {\it GEANT - Detector Description and Simulation Tool}, 
		CERN Program Library Long Writeup \textbf{W5013}, CERN (1993).

\bibitem{sibyll} Engel R. et al.,
		{\it Proc. $26^{th}$ Int. Cosmic Ray Conf.} Salt Lake City (USA)
		\textbf{1} (1999) 415; \\
		Engel, J. et al., {\it Phys. Rev. D} \textbf{46} (1992) 5013. 

\bibitem{muon} Antoni T. et al. - KASCADE-Collaboration, 
		{\it Astropart. Phys.} \textbf{16} (2002) 373. 

\bibitem{eastop} Aglietta  T. et al. - EAS-TOP-Collaboration, 
		{\it Astropart. Phys.} \textbf{21} (2004) 583. 

\bibitem{milke} Milke J. et al. - KASCADE-Collaboration, 
    {\it Proc. $27^{th}$ Int. Cosmic Ray Conf.} Hamburg
		\textbf{1} (2001) 241.

\bibitem{watson} Watson A.A., Proc. 25th Int. Cosmic Ray Conf. 
(Durban) Invited, Rapporteur, and Highlight papers, p 257 (1997).

\bibitem{epos} Werner K., Liu F.M. and Pierog T., {\it Phys. Rev. C} 
		\textbf{74} (2006) 044902.

\bibitem{akeno} M. Nagano et al., {\it J. Phys. G: Nucl. Part. Phys.} 
		\textbf{10} (1984) 1295.

\bibitem{alvarez} Alvarez-Muniz J. et al.,
		{\it Astropart. Phys.} \textbf{27} (2007) 271. 

\end{thebibliography}
\end{document}